\documentclass[12pt]{iopart}

 \usepackage{bm}
\usepackage{graphicx}
\usepackage[backref]{hyperref}
\usepackage{natbib}
\usepackage{units}
\usepackage{color}
\usepackage{nicefrac}

\DeclareTextFontCommand{\emph}{\em}

\begin{document}

\title{Web-based description of the space radiation environment using the  Bethe-Bloch model}

\author{Emanuele Cazzola$^\dag$, Stijn Calders$^\S$ , Giovanni Lapenta$^\dag$}
\address{$^\dag$ Centre for mathematical Plasma-Astrophysics - KULeuven, Leuven, Belgium}
\address{$^\S$ Belgian Institute for Space Aeronomy (BIRA-IASB), Brussels, Belgium}



\vspace{10pt}


\begin{abstract}
Space weather is a rapidly growing area not only in scientific and engineering applications but also in physics education and in the interest of the public. 
We focus especially on space radiation and its impact on space exploration. The topic is highly interdisciplinary bringing together fundamental concepts of nuclear physics 
with aspects of radiation protection and space science. We present a new approach to presenting the topic by developing a web-based tool that combines some of the fundamental concepts from these two fields in a single tool that can be developed in the context of advanced secondary or undergraduate university education. 

We present DREADCode,  an outreach or teaching  tool to asses rapidly the current conditions of the radiation field in space. 
DREADCode uses the available data feeds from a number of ongoing space missions (ACE, GOES-13, GOES-15) to produce a first order approximation of the dose an astronaut would receive during a mission of 
exploration in deep space (i.e. far from the Earth's shielding magnetic field and from the radiation belts). 

DREADcode is based on a easy to use GUI interface available online from the European Space Weather Portal (\url{http://www.spaceweather.eu/dreadcode}). The core of the radiation transport computation to produce the radiation dose from the observed fluence of radiation observed by the spacecraft fleet considered is based on a relatively simple approximation: the Bethe-Block equation.  DREADCode assumes also a simplified  geometry and material configuration for the shields used to compute the dose. The approach is  approximate and it sacrifices some important physics on the altar of a rapid execution time allowing a real time operation scenario.

There is no intention here to produce an operational tool for use in the space science and engineering. Rather we present an educational tool at undergraduate level
that uses modern web-based and programming methods to learn some of the most important concepts in the application of radiation protection to space weather problems.

\end{abstract}

%
%
%
%
%

\section{Introduction}

Exploring space is no gala dinner. 

{\color{black}
The space environment is particularly rich in ionizing radiations and can be highly lethal~\citep{schimmerling2003radiation}. 

Radiation encompasses all forms of energy emitted by a specific source moving through space by means of either particles or electromagnetic waves.
More specifically,
 ionizing radiation refers to the  strong interaction between high energetic charged particles and the material they travel into. In turn, the ionizing radiation is further split in two categories: directly and indirectly 
ionizing radiation. The former considers all the charged particles directly coming from the source and not changing its charge status on the way. The second group instead consists of the secondary 
ionizing particles produced \emph{after} the interaction with a primary particle, which can be either charged or neutral, with the medium.

Two main sources of ionizing radiation are present in the interplanetary space: cosmic rays and solar energetic particles (SEP)\citep{bothmer2007introduction}. Both vary in time. Cosmic rays are always present at levels that when accumulated over time pose serious risk of cancer to people and of damage to technology. 
SEP are sporadic but when they hit, the damage can be devastating. 

Solar energeric particles are released during specific solar events, also called solar eruptions, including coronal mass ejections and solar flares, which are able to send out an enormous amount
of highly energetic particles all over the heliosphere. SPE also includes those particles eventually accelerated through the interaction with some specific event in space, such as shocks.

}

All these threats are part of the rapidly growing discipline named by space weather, including all astrophysical and space processes impacting humans and technology in space or on the Earth 
(see \cite{schrijver2015} for more information and a complete definition of space weather).

A strong solar eruption can produce dangerous doses of radiation even at relatively low orbits where the geomagnetic field provides some shielding and definitely lethal for deep space missions.   Manned missions of exploration  have to be planned with careful attention to the space weather threat posed by radiation.  For this task numerous statistical studies  and computational tools have been developed so that a mission is planned under different scenarios and considering the plausible worst case scenario. 
 de
Models include solar energetic particles produced by space weather events \citep{king,Feynman1,Feynman2,Feynman3,ESP1,ESP2,ESP3,Rosenqvist,Nymmik_SPE},  galactic cosmic rays \citep{nymmik1992,nymmik1996,ISO15390}, as well as the radiation belts environment \citep{AP8,AE8,Vette1991}.

Many tools have been proposed in literature in order to evaluate the radiation
risks that astronauts undergo during space missions \citep{SPENVIS,SEPEM,singleterry2011,slaba2010,singleterry2011,schwadron2010,CREME96,OMERE}. 

The focus is of course on the near Earth environment where for decades now human presence has been limited to. But the tools also include deeper space exploration. Once a deep space mission is ongoing, a task unfortunately not occurred since Apollo 17 in December 1972, 
there is another need: to monitor space radiation in real time to assess the current conditions and make predictions of developing solar events. 
This task is not of immediate need because of the lack of manned missions since 1972, but is nevertheless a task to consider in planning a future return to manned exploration. World-wide space agencies and private ventures are now planning to undertake interplanetary journeys to  celestial bodies,
such as the Moon, Mars and asteroids \citep{reichert2001,NASA_exploration,NASA_report,NASA_objectives}. Radiation risks during these missions will be extreme \citep{ESA_HUMEX_report,mars-trip,mars-trip1,Norma,Durante-Cucinotta,facius}.

The study of this topic requires fundamental concepts of nuclear physics that determine how radiation interacts with matter. But understanding the topic requires one to consider the different sources of high energy particles in space and aspects typical of radiation protection to assess the impact on technology or on astronauts.
 
We present here an educational project that uses the existing satellites monitoring space (and future as they will become available)  to asses quickly, albeit approximately, the current risk of dose exposure.  
The concept of dose and the related different ways to compute are described in the section \ref{sec:dose}.

The goal is for a real time tool giving the current conditions available from the feeds of the existing monitoring satellites.   The  focus is only on deep space missions (i.e. we are not concerned here with radiation belts where other tools are already available 
\citep{AP8,AE8,Vette1991}). 

The idea of the present work, then, is to develop a new fast tool
able to assess doses received by human tissues taking into account
particle datasets directly recorded by satellites, in order to provide
the most realistic assessment of the present space situation. 
Furthermore, to reduce the computational effort,
as core of the computation, a first-order non-linear differential
equation is considered {\it in lieu }of Monte Carlo simulations. 

This design choice make the computation faster and more
general, although the final results will be less precise and less
specific compared with those obtained from Monte Carlo. 

{\color{black}

Nevertheless, this approach still maintains a good approximation in light of the physics carried by heavy charged particles. It is well estalished that a single charged particle behaves quite more differently
than a neutral particle in passing through the matter. We know that the mean free path will be much shorter, and the interaction predominantly due a long-range 
electromagnetic interaction with the target atoms, making the single collisions negligible. As result, the overall trajectory within the matter turns out to appear in first approximation straight,
removing the necessity of using a Monte Carlo random-walk.
Moreover, a set of particles with the same properties will behave nearly in the same way, being not affected by the single point-to-point interaction.

}

Based upon these latter considerations,  the macroscopic quantity  \emph{stopping power} is defined as the capability of a particular material to slow down an incoming particle through 
different type of interactions. The result is the loss of the initial particle energy along its path through the target material, so that the stopping power is  related to the concept of
\emph{linear energy transfer}. The practical use of this quantity is provided in section \ref{sec:bethe-bloch}, where the Bethe-Bloch equation is introduced.

Given the educational goal,  at the undergraduate level, this approach, even with its limitations for real world professional applications,  is preferable as a way to  make students 
familiar with real-time monitoring of space radiation and the assessment of its dangerousness.
Additional educational endeavours can ask the students to replace the simple transport model used here with more advanced and accurate radiation transport methods.

The tool presented here, given the topic of application, has been named DREADCode. In summary its peculiarities are:

\begin{itemize}
 \item data values directly taken from satellites, which leads to an almost real-time assessment and to a more realistic approach
 \item fast computation of radiation transport thanks to the use of the Bethe-Bloch equation considering all the correction factors introduced over time
 \item opportunity to choose material and composition of each layer is supposed to shield the incoming particles
 \item opportunity to assess the dose situation at different distances from the Sun, which means considering the Interplanetary Space surrounding those planets closer to the Sun. 
 Accuracy and precision at longer distances need further investigations, {\color{black} which are beyond the scope of this work and not well understood yet in the literature. Considering only the portion of the interplanetary 
 space we are interested (i.e. utmost up to the Mars distance) allows for removing the constraint concerning its high spatial variability. Temporal variations are instead intrinsically well described
 by the input satellite dataset.}
\end{itemize}

The code is written in MATLAB for its educational scope and has been recently made available online on the European Space Weather Portal website \citep{ESWP}, within the \emph{biological effects} section \citep{DREADCode}. 

{\color{black}
The paper is organized as follows. Given the highly interdisciplinary nature of the material presented, section \ref{sec:Rad_source} gives a brief description of
the radiation sources of major importance in space, as well as an introductory
on the main dosimetric quantities further used in the work.
Section \ref{sec:DREADCode} describes the DREADCode structure, highlighting its modules and their integrations. Section \ref{sec:results} provides validation and verification results,
while conclusions and future directions are provided in section \ref{sec:conclusions}, together with the educational potentiality of the tool.
}


\section{Radiation Sources and Effects} \label{sec:Rad_source}

{\color{black}

During interplanetary missions, astronauts are going to face 
ionizing radiations coming from different sources: besides the aforementioned sources (i.e. SEP and Cosmic Rays), it is worthy to additionally mention the
solar wind and the radiation belts.

The former represents the continuous stream of low energy charged particles from the Sun (predominantly protons at nearly $400\ \unit{\frac{km}{s}}$
through $800\ \unit{{km}{s}}$, meaning few $\unit{KeV}$ or less) in a way similar to the terrestial wind. For our purposes, the solar wind can be neglected as modern spacecraft are already equipped with
properly designed shields against these particles.

The radiaton belts are instead regions of relatively high energy particles trapped within the planetary magnetic fields, and literally surrounding those planets provided of a magnetosphere
(e.g. Earth and Mercury). Similarly to the solar wind, also this source of radiation can be neglected due to the short time spent by astronauts within them during the mission.

}

In contrast, we cannot neglect the solar energetic particles at all, 
as being very unpredictable, either in time or in magnitude,
and very problematic for interplanetary journeys. 
{\color{black}
To date, it seems in fact to be impossible
to properly design shields  without increasing dramatically masses,
volumes and, consequently, mission costs.
These particles are directly originated in the Sun and released after the occurrence of extreme solar events, such Coronal Mass Ejections (CMEs).
}

Finally, Galactic Cosmic Rays and Anomalous Cosmic Rays are fluxes
of particles, ranging from proton to uranium nucleii, coming from
galactic and extra-galactic origins, whose path is undefined as they
can undertake many deflections within the {\it interplanetary magnetic
field} (IMF) {\color{black} before reaching the target}. Consequently, the population of these particles inside the
{\it heliosphere} is strictly function of the IMF strength, which
in turn is function of the solar activity. Although we considered
GCR and ACR as the same entity, indeed, the sources of these particles
are strongly different \citep{durante2011}. GCRs are basically charged particles
coming from the deep space, flowing around the {\it interplanetary
magnetic field} lines and being accelerated when eventually encountering particular astrophysical events, such as
shocks. ACRs are instead neutral
particles coming from {\it interstellar material}, which happen to be
weakly ionized through the interaction with the Solar Wind as soon as they enter the {\it heliosphere} (\citet{schrijver2010heliophysics}).
The latter are therefore mainly composed by {\it weakly-ionized}
heavy nucleii which have lost only
their outer atomic electrons, resulting much less influenced
by the IMF than GCRs. Additionally, they show a significantly lower energy spectrum than
GCRs. 



\begin{figure}
\begin{centering}

 \includegraphics[scale=0.25]{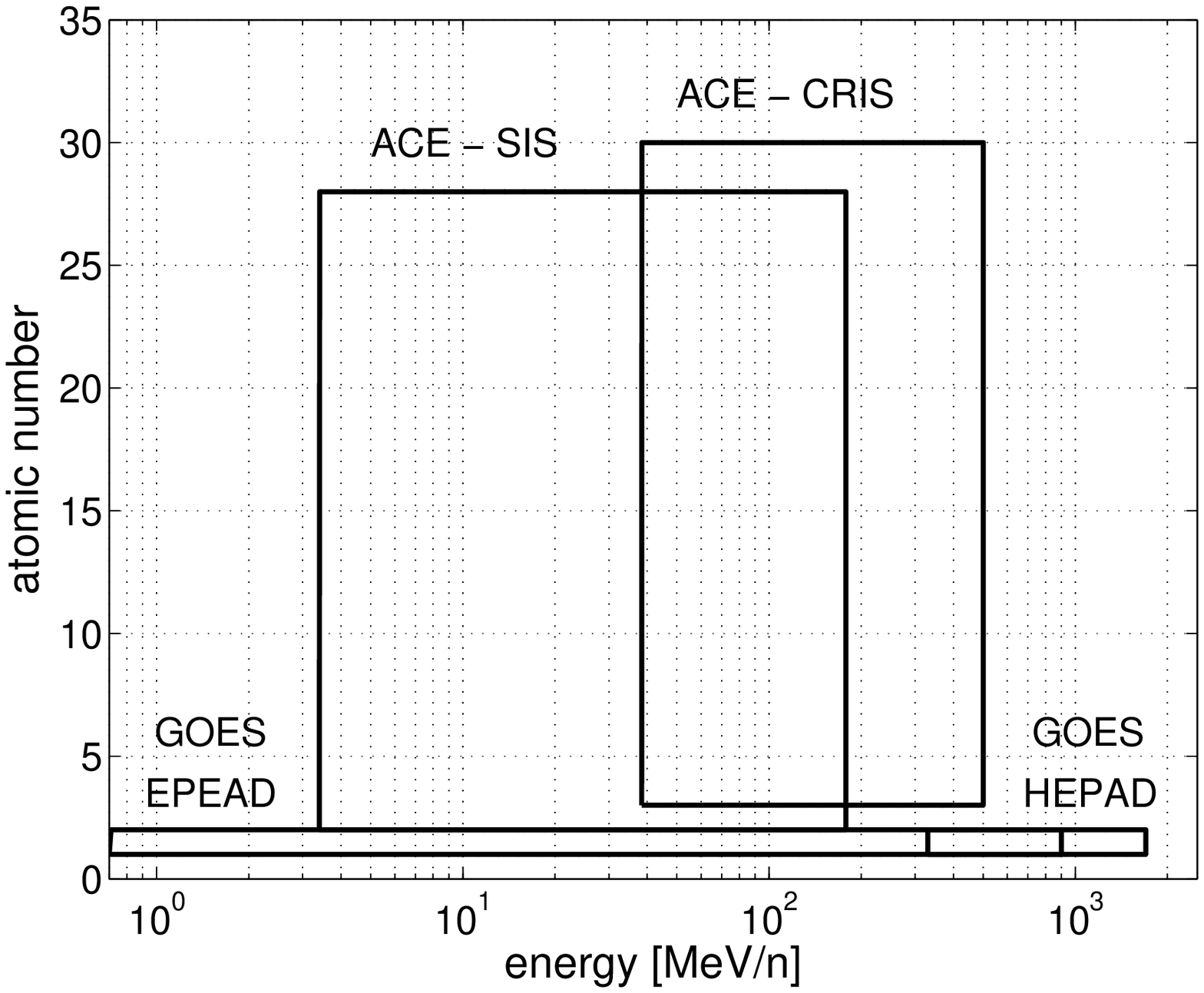}
 \includegraphics[scale=0.25]{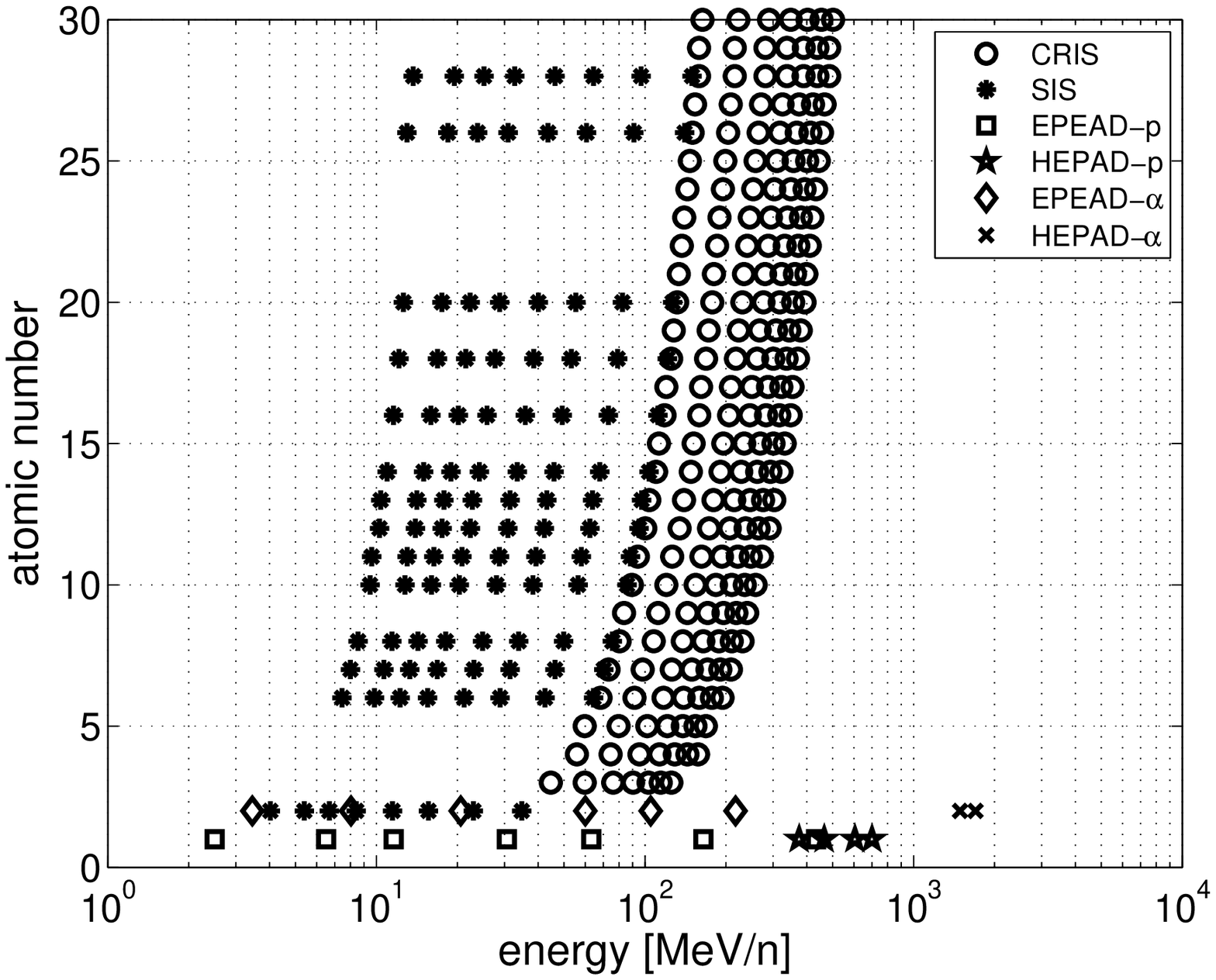}
 \includegraphics[scale=0.25]{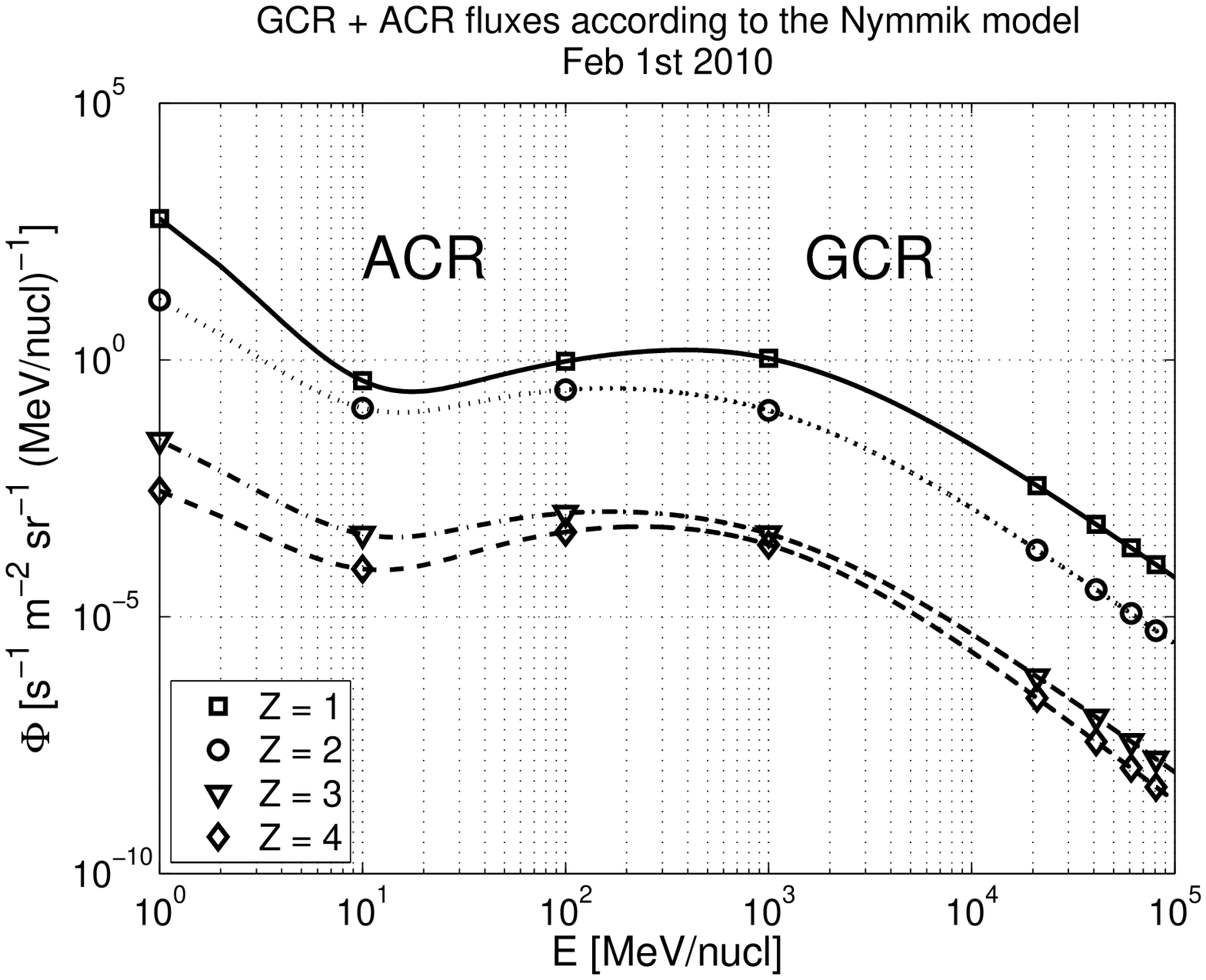}
 \label{fig:data_sat}
 \caption{\textcolor{black}{Spectra of data recorded by the satellites considered within DREADCode. The left panel represents 
 an uniform and continuous spectrum for each ion, while the middle panel 
 points out the more realistic average energy values measured for each energy bin and for each ion taken into account. The right panel gives an example of particles fluxes 
 computed with the Nymmik's model. }.}
 
 \end{centering}
\end{figure}

\subsection{Dose and Biological Effects} \label{sec:dose}

{\color{black}
Main concern of radiations is the danger they can cause to devices and, mostly, to biological tissues. 
Historically, the main difficulty in assessing their dangerousness has always been finding the best approach to link the macroscopic physical causes, 
such as the presence of a large scale radiation field, to the microscopic effects caused at lower scales inside the matter.
An ultimate common agreement on what is the quantity best accomplishing this task is still far to come.
Even though being no the solely suggested quantities, DREADCode takes into account the currently most used standard definitions for radiometry and dosimetry. 
These definitions are scientifically motivated and are incorporated into regulatory assessments \citep{ICRP2007gloss,protection2007,protection2010}.
}


%
%
%
%
%
%
%

\begin{table}
\begin{centering}

\begin{tabular}{ | p{.75\columnwidth}| p{.15\columnwidth}| }
\hline 
\textbf{Radiation $R$} & \textbf{$w_{R}$}\tabularnewline
\hline 
\hline 
photons & $1$\tabularnewline
\hline 
electrons and muons & $1$\tabularnewline
\hline 
protons and charged pions & $2$\tabularnewline
\hline 
$\alpha$ particles, fission fragments and heavy ions & $20$\tabularnewline
\hline 
neutrons & \emph{specific function} \tabularnewline
\hline 

\hline 
\textbf{Tissue $T$} & \textbf{$w_{T}$}\tabularnewline
\hline 
\hline 
bone-marrow, colon, lung, stomac, breast, \emph{remainder tissues} & $0.12$\tabularnewline
\hline 
gonads & $0.08$\tabularnewline
\hline 
bladder, oesophagus, liver, thyroid & $0.04$\tabularnewline
\hline 
bone surfce, brain, salivary glands, skin & $0.01$\tabularnewline
\hline 
$\sum\limits_{T} w_{T}$ & $\sim1$\tabularnewline
\hline 

\end{tabular}
\par\end{centering}

\caption{\label{tab:Radiation-weighting-factor}Radiation weighting factors $w_{R}$ and Tissue weighting factors $w_{T}$
for the most common radiation types and organs. For neutrons, a {\it specific continuous function} is suggested in \citet{protection2007}.}

\end{table}

\paragraph{Effective Dose and Ambient Dose Equivalent}

{\color{black}

Main output of the code are the effective dose and the ambient dose equivalent.

The former is agreed to be the best approach to directly connect 
the microscopic biological effects to the macroscopic radiation fields in which the target is eventually found. 
In particular, it has been proved that different
organs or tissues of the human body have different reactions to the same incoming radiation. This property is called {\it radio-sensitivity}.
The radio-sensitivity is intimately
related to both the cellular reparation capability of the single organ or tissue,
and the cellular renovation rate \citep{shultis}. 
The effective dose gives an estimation of the different radio-sensitivity of a biological tissue and is computed as

\[
E=\sum_{T}w_{T}\cdot H_{T}\ \unit{\unit{\left[Sv\right]}}
\]

\[
\sum_{T}w_{T}=1
\]
where $H_{T}$ is the {\it equivalent dose }received by the tissue
$T$ and $w_{T}$ are coefficients called {\it tissue weighting
factors}. The total sum of these coefficients is equal to 1 for a
{\it whole body} exposure. Recent values of $w_{T}$ are given in table \ref{tab:Radiation-weighting-factor}, together with the radiation weighting factors necessary to compute 
the equivalent dose. In analyzing these values, which are continuously updated
through experimental and simulation campaigns, noticeable is the relation between the greatest
factors and those organs expected to have higher cellular renovation rates.
Like the equivalent dose, the effective dose is 
expressed in $\unit{Sv}$ - {\it Sievert}.

Practically, the effective dose turns out to be very helpful when the exposure
is well known and defined. However, when it comes to evaluate the ionizing effects
in cases where the exposure of the target 
is not totally clear, the latter becomes less handy. 
Glaring example is the variable position astronauts are keeping during the different stages of a long-term mission.

A more suited dose quantity has therefore been 
introduced, namely the {\it ambient dose equivalent}. According to the definition appointed in \citet{ICRP2007gloss} and \citet{protection2007,protection2010}, this quantity represents the \emph{equivalent dose} 
computed at $10\ \unit{mm}$ radially inside a \emph{ICRU-sphere}, when the radiaton field is intended to be aligned (i.e. oriented), expanded and oppositely oriented with respect to the sphere, as indicatively
shown in figure \ref{fig:ICRU_sp}.
The \emph{ICRU sphere} is a $30\ \unit{cm}$-diameter
sphere made of specific elements ($76.2\ \%$ O, $11.1\ \%$ C,
$10.1\ \%$ H and $2.6\ \%$ N) recalling the human tissue composition, with density equals to $1\ \nicefrac{g}{cc}$ \citep{ICRUreport33}.

\begin{figure}
 \begin{centering}
  \includegraphics[scale=0.40]{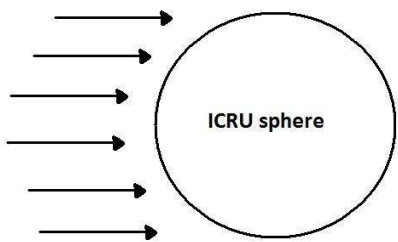}
 \par\end{centering}

 \caption{\label{fig:ICRU_sp}Sketch representing the aligned (i.e. oriented) and expanded radiation field around the \emph{ICRU-sphere}, according to the definition \citet{protection2007,protection2010}.}
 
\end{figure}

}

\section{DREADCode: main features} \label{sec:DREADCode}
\begin{figure}
 \begin{centering}
  \includegraphics[scale=0.9]{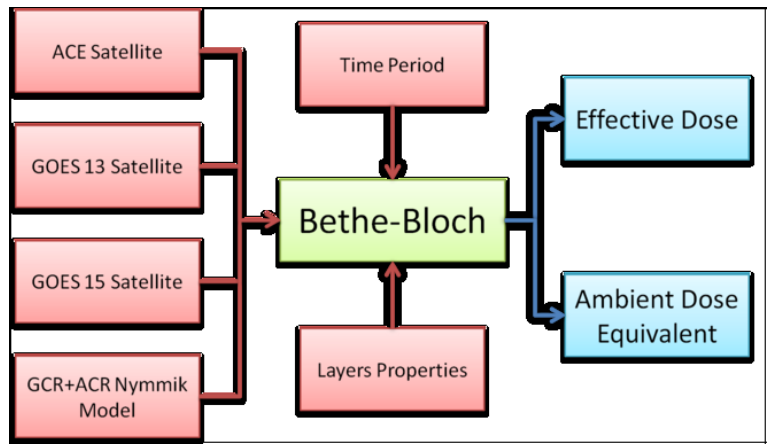}
 \par\end{centering}
 
 \caption{\label{fig:code_cartoon}Flowchart summarizing the main DREADCode's inputs and outputs.}

\end{figure}
DREADCode has a modular structure.  Figure \ref{fig:code_cartoon} shows a flowchart describing the main inputs, the computational core and the main outputs.  
Besides data sources, time period and shield properties, the user can also set up other features, including the exposure and the scaling 
 factor, as well as what specific coefficients one prefers using for each output assessment.

The main features of the DREADCode are listed in order. 

First, as input DREADCode  receives particle flux data directly from satellites
recordings.  Only the true sources of radiation concern are considered: charged nuclei, from
{\it proton }to {\it uranium }nuclei. Future versions will consider also neutral particles (neutrons, gamma rays), for now neglected.

This approach leads to an (almost) real-time assessment. A delay is still present due to the raw data satellite's processing required before the data is made available by the responsible agency (NOAA for the data sources used). 
Moreover, the user has to insert the shields or layers properties, such as thickness, material and elements composition.

Second, a radiation transport calculation is  required to evaluate how many particles with
a particular energy are able to penetrate the shields under consideration. For this task, a simplified model is used, instead of the more computational demanding  
Monte Carlo method. This aspect will be subject to future evolution and improvement and is currently motivated by the need for computational speed to achieve real-time performance.

\begin{figure}
\begin{center}
 \includegraphics[scale=0.6]{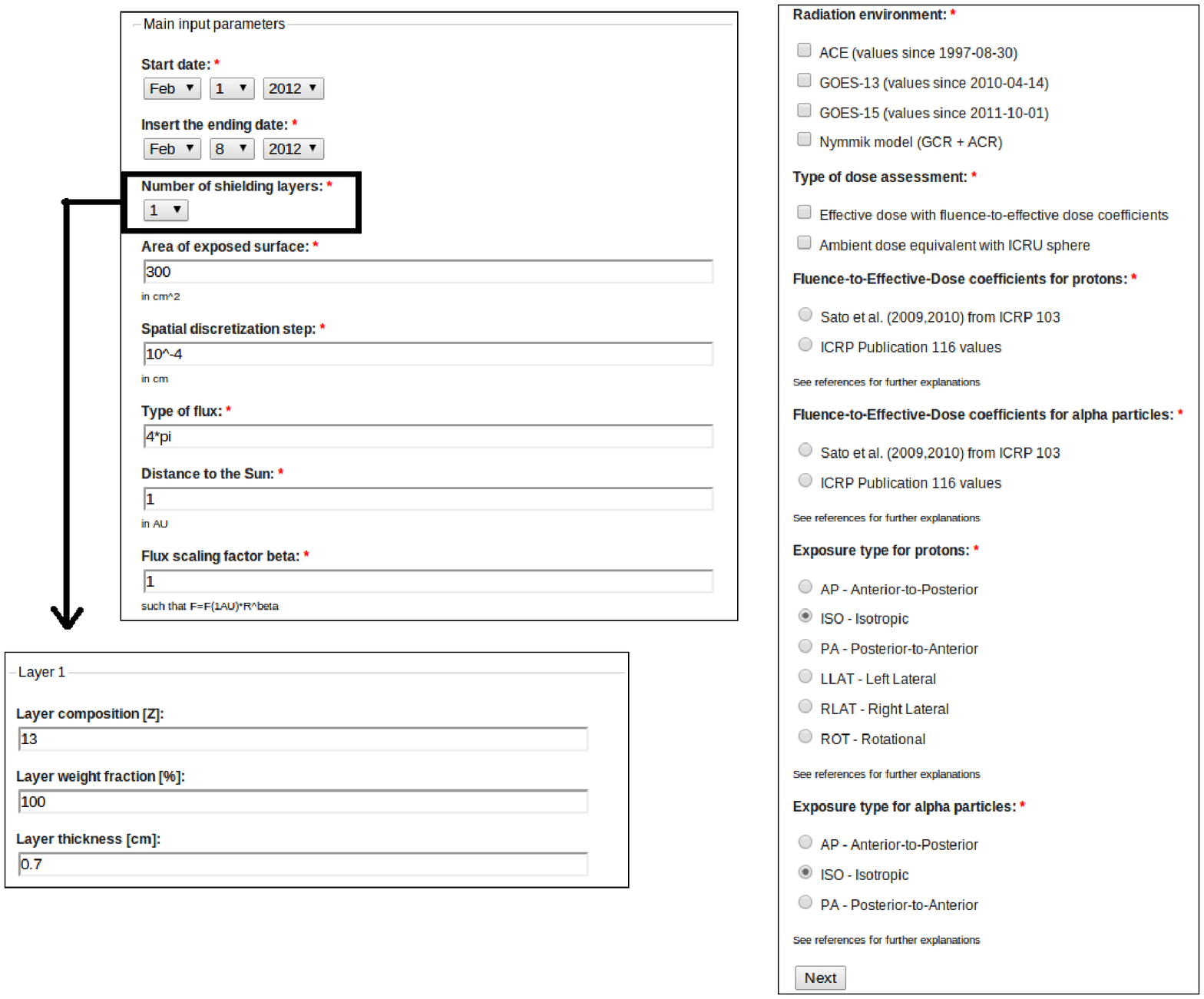}
 \caption{\textcolor{black}{Set of snapshots directly taken from the on-line version of DREADCode, hosted on the European Space Weather Portal. Here it is possible to notice three main sections: 
 inside the \emph{main input parameters} one can insert the time period and other general inputs concerning the exposure. In the middle, instead, it is shown the layer properties mask, where the user
 defines all those layers existing between sources and the target. Finally, on the right the \emph{radiation environment} section is devoted to the radiation source choice and the output features}}.
  \label{fig:inputGUI}

 \end{center}

\end{figure}

\textcolor{black}{An example of what stated above is given in figure \ref{fig:inputGUI}, where snapshots taken from the on-line DREADCode are shown. In particular, after inserting the general necessary 
inputs, such as the time period, distance from the Sun and type of exposure of the analyzing object, the user has also to define the layers properties. Firstly, the knowledge of how many layers are
present between the external radiation sources and the final target, i.e. human tissues, it is necessary. This figure represents the case with a single layer, e.g. an \emph{aluminium}-based wall. 
Thereafter, for each layer one has to define its material, which may be single or multiple compounds, the relative weight fraction composition, in percentage, and, finally, its thickness}. 

\textcolor{black}{At the end, the user can select what type of particle data input considering, i.e. which satellite and/or model, as well as the desired output. In case of the \emph{effective dose}, 
one has also to  select the type of exposure and what conversion coefficients to use}.

\begin{figure}
 \begin{center}
  
  \includegraphics[scale=0.7]{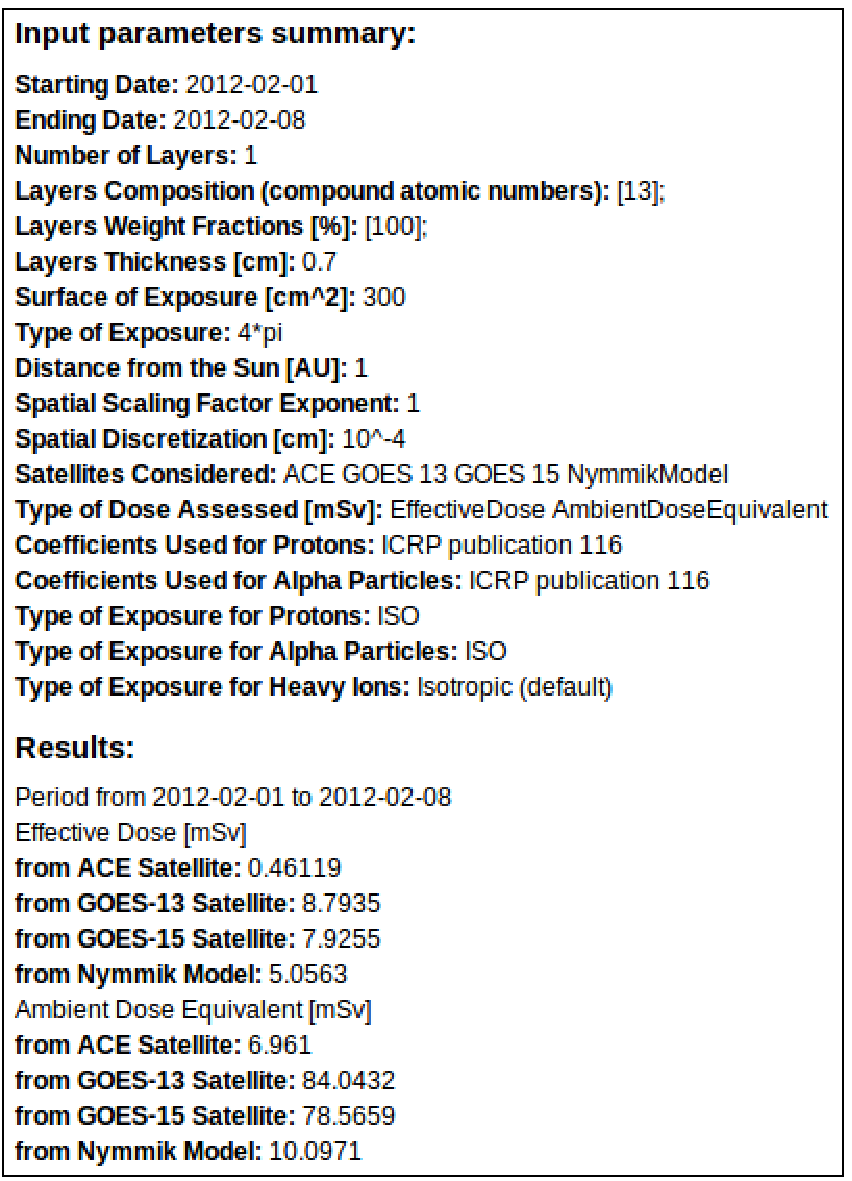}
  \caption{\textcolor{black}{Snapshot  of the DREADCode's on-line version representing the final page with results. This mask shows a summary of all input parameters inserted, together with results for  those 
  requested outputs}.}   \label{fig:outputGUI}

 \end{center}

\end{figure}

\textcolor{black}{Results are finally represented with a graphics interface page. An example is given in figure \ref{fig:outputGUI} showing the case analyzed later in section \ref{sec:results}. 
This page gives a summary of the input parameters, as well as results for any requested output}.

\subsection{Inputs of the Code}

The code receives the following main inputs:
\begin{itemize}
\item the temporal period, i.e. the {\it starting date }and the {\it ending
date}. These values should not exceed the ranges pointed out by satellite's
websites
\item the satellite(s) data source, including ACE, GOES-13 and GOES-15.
The Nymmik's models for GCR and ACR fluxes \citep{nymmik1992,nymmik1996,ISO15390}
have been also implemented to compare model and data-driven results, \textcolor{black}{as well as use them when satellites data are not available, i.e. future periods or maintenance}
\item the number, composition, weight fractions for each compound and thicknesses
of layers (shields) between the incoming particles and the target
\item the total exposure surface of the target, such as the external spacecraft wall, the space suit surface or the human body surface. Examples of models to compute 
the human body surface are pointed out in \citet{verbraecken2006} and in \citet{yu2003}.
\item the angular exposure, in order to integrate the flux values overall the suitable solid angle
\item the distance $R$ of the target from the Sun, expressed in $\unit{AU}$
\item the {\it exponent scaling factor} $\beta$, which properly scales the flux/fluence
values spatially from $1\ \unit{AU}$ (location where these values
are supposed to be recorded) to distance $R$ from the Sun, such that

\[
 F(R) = F ( 1\ \unit{AU} ) \cdot R^\beta
\]

This value has been kept as free parameter according to the uncertainty on the spatial scaling stated in \citet{smart2003}.

\item the output's options and the suitable coefficients to use for the
computation
\end{itemize}

Concluding on the main code features, it is remarkable that any time the user inserts an erroneous value inside the GUI, the code returns an output page indicating what the insertion error might be. This also occurs when the code
itself is not able to link with the necessary websites or datasets. The most common fails derive from an incorrect date insertion, since any implemented satellite has a different temporal range to consider. 
Hence, the user should already
be informed about the properties of the considered satellites. Particularly, the necessary starting date limit have been pointed out in the GUI, while the limited ending date obviously changes as function of the satellite.
Further error indications have been configured for other common insertion mistakes and network connection failures.

\subsection{Core of the Code} \label{sec:bethe-bloch}

The core of the code is the Bethe-Bloch's equation,
which returns the value of the {\it stopping power }for particles
with particular energy inside a prior-defined medium. The  {\it stopping
power }$S$ is an useful physical quantity since heavy
charged particles normally travel along straight lines within a medium,
 releasing great amounts of energy in a very short path, due to their
great mass and charge. 

\begin{figure}
\begin{centering}
\includegraphics[scale=0.6]{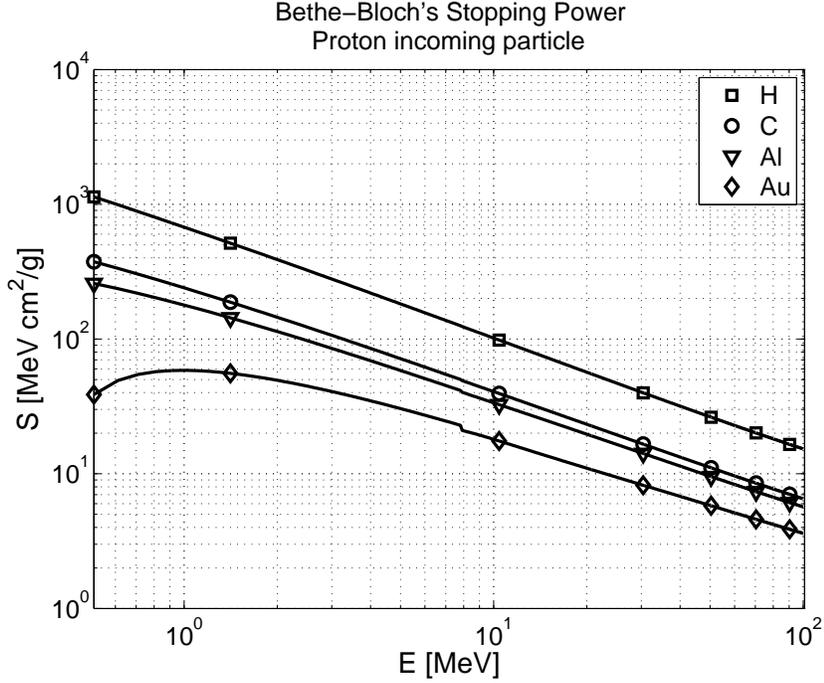}

\par\end{centering}

\caption{\label{fig:Stopping-power's-profile}Stopping power profile as function
of energy for incoming proton and some target, computed from equation \ref{eq:bethe-bloch}.}

\end{figure}

The Bethe-Bloch's equation is the following

\begin{equation}
S=-\frac{dE}{dx}=Kz^{2}\frac{Z}{A}\frac{1}{\beta}\left[\frac{1}{2}\ln\frac{2m_{e}c^{2}\beta^{2}\gamma^{2}T_{max}}{I^{2}}-\beta^{2}-\frac{C}{Z}-\frac{\delta}{2}\right]\label{eq:bethe-bloch}
\end{equation}
\[
T_{max}=\frac{2m_{e}c^{2}\beta^{2}\gamma^{2}}{1+2\gamma\frac{m_{e}}{M}+\left(\frac{m_{e}}{M}\right)^{2}}
\]
where 

$K=4\pi N_{A}r_{e}^{2}m_{e}c^{2}$, such that $\frac{4\pi N_{A}r_{e}^{2}m_{e}c^{2}}{A}=0.307075\ \unit{MeV}$
for $A=1\ \nicefrac{g}{mol}$

$N_{A}=$ Avogadro number

$r_{e}=$ electron radius

$m_{e}=$ electron mass

$Z=$ atomic number of the \emph{target}

$A=$ atomic mass of the \emph{target}

$z=$ incident particle charge, in unit $e$

$\beta=\frac{v}{c}$

$\gamma=\frac{1}{\sqrt{1-\beta^{2}}}$

$\delta=$ density effect correction parameter

$C=$ shell correction parameter

$I=$ mean excitation potential of the {\it target}

$T_{max}=$ maximum transferable energy to an electron after a collision\\

The {\it mean excitation potential} $I$ is computed according to the model 
suggested by \citet{sternheimer}, who charted these
values for several elements.

When the incoming charged particles have high energies, the electric field associated to them also increases, which influences the stopping power (the  $\beta \cdot \gamma$ term inside the logarithm would increase). 
Indeed, real media are already able to limit these effects, therefore the {\it density effect correction }$\delta$ has been introduced to correct the stopping power at high energies \citep{fermi}. In
order to consider this effect we again use  the model suggested by \citet{sternheimer}.

Conversely, the {\it shell correction }$C$ corrects the
error of the equation associated to low energies.
The model used here is shown and explained in \citet{barkas}.

Finally, notice that inside equation \ref{eq:bethe-bloch} there are
terms related to the target material, such as $A$,
$Z$, $\rho$ and $I$, and terms related to the incident particle properties,
such as $z$ and, indirectly, $M$. The unit of the  {\it stopping
power} is $\nicefrac{MeV}{cm}$, whether the mass density of the
medium is included, or $\nicefrac{MeV\ cm^{2}}{g}$ whether one
wants to remove the dependence of the target's mass density from the computation.

To illustrate the performance of the Bethe-Bloch approach, Figure \ref{fig:Stopping-power's-profile}
shows the {\it stopping power} variation as function of energy for incoming protons in different media. The minimum cut-off energy is $0.511\ \unit{MeV}$, which is the value \textcolor{black}{normally} 
associated to electron bound energy.
Results obtained from equation \ref{eq:bethe-bloch} have been compared with those obtained from other availale tools both in literature and on-line, such as Nucleonica \citep{nucleonica}, 
the ASTAR-PSTAR of the NIST databank \citep{NIST} and SRIM \citep{SRIM,ziegler2010srim}. Results are shown in table \ref{tab:stoppingTable}. Clearly, while approximated, the approach is sufficiently accurate for a first order real time assessment.

\begin{table}
 \begin{centering}
 
 \begin{tabular}{|c|c|c|c|c|}

 \hline
 
 \multicolumn{5}{|c|}{\textbf{Stopping Power} $\left[ \unit{\frac{MeV\ cm^{2}}{g}} \right]$} \\ \hline
  
  \hline 
\textbf{Energy} $\mathbf{\left[ \unit{MeV} \right]}$ & \textbf{Nucleonica} & \textbf{NIST} & \textbf{SRIM} & \textbf{DREADCode} \tabularnewline
\hline 
\hline 
$0.2$ & $3.735 \cdot 10^{2}$ & $3.715 \cdot 10^{2}$ & $3.730 \cdot 10^{2}$ & $3.347 \cdot 10^{2}$ \tabularnewline
\hline 

$2$ & $1.109 \cdot 10^{2}$ & $1.095 \cdot 10^{2}$ & $1.108 \cdot 10^{2}$ & $1.136 \cdot 10^{2}$ \tabularnewline
\hline 

$10$ & $3.398 \cdot 10^{1}$ & $3.376 \cdot 10^{1}$ & $3.396 \cdot 10^{1}$ & $3.362 \cdot 10^{1}$ \tabularnewline
\hline 

$100$ & $5.691 $ & $5.678$ & $5.689$ & $5.677$ \tabularnewline
\hline 

 \end{tabular}

 \end{centering}
\caption{\textcolor{black}{Comparison of the stopping power computed with different tools, including the Bethe-Bloch model described in this manuscript}.} \label{tab:stoppingTable}
\end{table}

%
%
%
%

\subsubsection{Bethe-Bloch equation for Compounds}

Equation \ref{eq:bethe-bloch} can also be used when the target
material is a mixture or multi-compounds medium by using the Bragg's
additivity rule (\citet{bragg_add,seltzer1982}), which approximates the target as a sequence of mono-component
layers by basically considering the whole stopping power as linearly
proportional to each component's stopping power, with mass fractions
$w_{j}$ such that

\begin{equation}
\left(\frac{dE}{dx}\right)_{comp}=\sum_{j}w_{j}\cdot\left(\frac{dE}{dx}\right)_{j}
\end{equation}

\[
w_{j}=\frac{n_{j}\cdot A_{j}}{\sum_{k}n_{k}\cdot A_{k}}
\]

\subsection{Outputs of the Code}

The code allows the user to assess two different doses: the \emph{effective
dose }and the {\it ambient dose equivalent}.

\subsubsection{Effective Dose Assessment}

After evaluating the fluence of particles able to cross the shields configured, it is possible to obtain the effective dose by multiplying the fluence by specific
coefficients. These coefficients, usually called {\it fluence-to-effective-dose-coefficients},
directly derive from experimental and accurate simulation campaigns
aimed to evaluate doses on human tissues inside specific radiation
environments. They are normally looked up as function of energy and
particles fluence, with the chance to interpolate when values do not
match. Many coefficients can be found in literature, but the code
considers those published in \citet{sato2009,sato2010,ICRP2010coeff}.

Beside the computation speed, this solution turns out particularly reliable thanks to the wide
studies devoted to compute these coefficients, which are continuously
updated and published in the official radio-protection reports \citep{ICRP2010coeff}.

\subsubsection{Ambient Dose Equivalent Assessment}

The user can also decide to evaluate the {\it ambient
dose equivalent} $H_{a}$. 

According to its definition, the code runs a further computation
by following the particles inside a prefixed compound recalling the {\it ICRU-sphere}, as long as they
achieve the $10\ \unit{mm}$ depth inside the
sphere. The computed energy release is, finally, multiplied by the proper factors pointed out in table \ref{tab:Radiation-weighting-factor} to assess
the ambient dose equivalent.

However, a computation of this kind would take more computational time than for the case of 
the effective dose assessment.

\section{Results and Discussion} \label{sec:results}
To validate and verify the tool, DREADCode was tested in several scenarios and compared with established tools available online.  
In this section, some results are shown. Since DREADCode
is a generic code to evaluate doses during interplanetary travels,
we decided not to follow any strictly mission schedule and test it for some different periods.

First of all, we are interested to compare results between a period
during which a solar event has been recorded and a period during which
solar events did not occur. Table \ref{tab:Comparison results}
compares these situations for the period February 1 - 8 (when no events were recorded) and period March 6 - 13 (when a solar event was recorded). These periods
have been selected according to the solar event list provided by NOAA \citep{NOAA_eventlist}.

\begin{table*}
\begin{centering}

\begin{tabular}{|c|c|c|c|c|}
\hline 
 & \multicolumn{4}{c|}{Feb 1st - Feb 8th 2012 (no solar events recorded)}\tabularnewline
\hline 
 & \textbf{ACE} & \textbf{GOES-13} & \textbf{GOES-15} & \textbf{Nymmik}\tabularnewline
\hline 
$E\ \unit{\left[mSv\right]}$ & $4.593\cdot10^{-1}$ & $8.7935$ & $7.9255$ & $5.0563$\tabularnewline
\hline 
$H_{a}\ \unit{\left[mSv\right]}$ & $6.961$ & $84.0432$ & $78.5659$ & $10.0971$\tabularnewline
\hline 
 & \multicolumn{4}{c|}{Mar 6th - Mar 13th (solar event recorded)}\tabularnewline
\hline 
 & \textbf{ACE} & \textbf{GOES-13} & \textbf{GOES-15} & \textbf{Nymmik}\tabularnewline
\hline 
$E\ \unit{\left[mSv\right]}$ & $4.0685\cdot10^{-2}$ & $70.092$ & $63.733$ & $5.0400$\tabularnewline
\hline 
$H_{a}\ \unit{\left[mSv\right]}$ & $4.1229$ & $1785.0703$ & $1644.9841$ & $10.062$\tabularnewline
\hline 
\end{tabular}

\par\end{centering}
\caption{\label{tab:Comparison results}Comparison between two different period
results: upper table summarizes doses when no solar event has been
recorded, whilst table below when solar event is occurred.}
\end{table*}

Values obtained from GOES satellites
seem to be coherent with the situation, since these satellites carry onboard
devices able to record both $H^{+}$ and $\alpha$ high energy particles,
even when solar events occur \citep{GOES_Data_Book}.
Unlike GOES, the ACE satellite turns out to be suitable to record
only GCR and high atomic number energetic particles, as well as solar
wind. In fact, the most important device for our purpose is the CRIS instrument, which measures nucleons ranging $3\leq Z\leq28$, i.e.
galactic cosmic rays. This device, however, does not properly cover 
periods of high solar activity, as it is usually switched off during solar events to prevent from damaging. This is the main reason why results
from ACE shows to be greater in the period February 1 - 8 than between March 6 - 13.

Comparing the results from the Nymmik's model with those from ACE,
it is possible to notice an agreement in the ambient dose
equivalent. This model is directly function of the IMF
strength, which in turn is function of the {\it monthly sunspots
number }and the solar activity. The number of the sunspots is directly
provided by the {\it Solar Influences Data Analysis Center } (SIDC)
of the {\it Royal Observatory of Belgium} (ROB) \citep{SIDC}.
This explains why the results are not influenced by the occurrence of solar events. On the other hand, a slight variation is noticed and explained with the
different solar activity in the two periods.
{\color{black}
As we expect, the second period, during which a solar event
was recorded, shows a nearly ten times higher dose value than
the case considering the GOES satellites.

We have finally compared DREADCode with similar tools already available online (namely \citet{SPENVIS} and \citet{SEPEM}) considering a mission to Mars according 
to the generic schedule pointed out in \citet{ESA_HUMEX_report}. Given the differences between the tools, assessment conditions have been kept nearly as  congruent as possible, even though some  
approximation to match these differences had to be done. Results are shown in table \ref{tab:results_comparison_conclusion}.
We notice that the effective doses show to be closely in agreement between all the tools.
Regarding the ambient dose equivalent, however, DREADCode appears to have a much more conservative approach than SPENVIS. SEPEM does not provide the assessment of this quantity.
The two computing methods are radically different. While DREADCode runs once again the same transport method into a prefixed layer resembling the ICRU-sphere, SPENVIS instead makes use of suited
coefficients similar to those used to compute the effective dose. Despite the results have the same order of magnitude, DREADCode seems to give a more conservative outcome.

}

\begin{table}
\begin{centering}

  \begin{tabular}{| c | c | c |} 
  \hline

  \multicolumn{3}{|c|}{\textbf{Mars Surface}} \tabularnewline
    
    \hline
    \hline
 
   \multicolumn{3}{|c|}{\textbf{Effective Dose $\mathbf{\left[ \unit{mSv} \right]}$}}
\tabularnewline
    
    \hline
    \hline

\textbf{SPENVIS} & \textbf{DREADCode}  & \textbf{SEPEM} \tabularnewline

\hline
1459 & 1066.088 & 1125.4   \tabularnewline
\hline
\hline

\multicolumn{3}{|c|}{\textbf{Ambient Dose Equivalent $\mathbf{\left[ \unit{mSv} \right]}$}}
\tabularnewline
    
    \hline
    \hline

\textbf{SPENVIS} & \textbf{DREADCode}  & \textbf{SEPEM} \tabularnewline

\hline
1170 & 3691.542 & - \tabularnewline
\hline

\end{tabular}

\end{centering}

\caption{\label{tab:results_comparison_conclusion}Comparison of results from SPENVIS, DREADCode and SEPEM.}

\end{table}

\section{Conclusions and Future Directions} \label{sec:conclusions}

{\color{black}

In conclusion, we presented the DREADCode, a handy tool to macroscopically assess the radiation doses on human tissues in a relatively short time, which considers a more realistic description of the existing
radiation field by directly using
satellites data. Conversely, similar and more advanced tools available to date only consider a generic view of the incoming radiation through statistical worst-case scenarios or semi-analytical models, which often turn out
to be too generic and unrealistic.
Instead, we believed that a significantly improvement of this kind of assessment could be reached by exploiting the wide fleet of satellites currently at disposal.
}
Moreover, unlike the statistical models based on the past worst-case events which only consider protons, in DREADCode the dose from heavy nucleii is also taken into account.

Also, being directly connected with satellites
allows us to consider an almost real time situation, with the only
delay due to the data time-processing. 

Parameter insertion is fast, easy and intuitive via a GUI,
and the computational costs are relatively low when compared
to more precise and accurate computations, such as those related to
Monte Carlo simulations.

Finally, the DREADCode is  meant to introduce the basic concept in a modern web-based frame. DREADCode is not intended to  address proper engineering shield
design,
{\color{black} given its high conservative approach. The main goal here in fact is to figure out what kind of particles is actually able to cross the shield configured by the user. No intention was given to the
shield design itself. }
Its generic profile, however,
allows the code to be used to assess the radiation shielding capacity for many general purposes, including spacecraft,
spacesuit and outpost shielding ability.

DREADCode is highly modular and flexible, allowing for further spin off educational endeavours that can  upgrade the code.  The following examples can  be taken into
consideration:
\begin{itemize}
\item Implementation of new satellite's datasets or new particles data
sources
\end{itemize}
It could be interesting to implement new and different particle data
sources, in order to improve the computation reliability on the long
period, since satellites have limited life-time, comparable to
few years, as well as to give more options to external users, such
as different data sources. 
\begin{itemize}
\item Inserting new coefficients for specific organs or tissues
\end{itemize}
So far, only the \emph{whole body exposure }has been taken into account.
Nevertheless, the user could want to compute doses for specific organs,
including the most problematic tissues. For this purpose, the literature
offers specific coefficients, such as those suggested by \citet{ICRP2010coeff,sato2010,sato2009,copeland,ferrari}.
\begin{itemize}
\item Considering secondary radiations
\end{itemize}

The physics can also be complicated at will. When charged particles interact with the medium, they can also generate
secondary radiations, including electrons and neutrons. While electrons
are normally not a big concern, neutrons may reveal themselves problematic
due to their high peak fluxes. This phenomenon turns out to be more
important when energetic particles from space interact with the {\it regolith
}of planetary surfaces, by increasing the neutron flux below the crust, whose
magnitude should be considered when long-permanence outposts are planned
to be built underneath the surface.

{\color{black}

Additionally, the tool presents itself as an useful educational tool connecting several problematic aspects typical of the interplanetary journeys.

We started addressing the idea of developing this tool after some experience acquired in covering the lectures on space weather recently proposed at KULeuven for undergraduate students.
This broad subject shows to be highly multi-disciplinary, including knowledge on astrophysics, material engineering and radioprotection.
While traditional lectures dealing with established terrestrial applications can count on significantly helpful hand-on sessions, courses on space weather rarely
allows students to get in touch with the daily practical issues faced by technicians in this field. 

In particular, we noticed the absence of a direct interconnection between the theoretical and observational approach on the study of astrophysical events and the effective causes led by them on the
human environment, with a particular gap noticed in the field of radioprotection in space. The latter is thought to be mainly caused by the absence of manned space missions over the last decades, which brought 
the education to heading different directions, such as the impact of radiation on human technologies instead of biological materials.
DREADCode has therefore been devised to fill this educational gap.

By using of the Bethe-Bloch's equation, undergraduate students additionally learn more about the radiation-matter interaction with highly energetic charged particles, as well as the difficulty to find the
proper materials able to suitable shielding the radiation without increasing masses and costs.
Moreover, they are going to tackle the high uncertainty in setting realistically the target conditions, given the wide variability of the possible scenarios during such long missions.

Finally, students become familiar with the elevated order of magnitude of dose reached in space, which are very rarely met in terrestrial conditions, and the latter surely in accidental situations. 
These values
pose further issues in evaluating the macroscopic biological effects on human tissues. In fact, whether the typical concerns of terrestial radiation fields are mainly focused on long-term low-doses cancer-risk 
expositions, radiations in space bring out a new synergistic combination of constant low-level doses from GCRs and sudden acute exposures in case of solar events, leading to a multi-scale phenomenon
rarely found in terrestrial applications.

}

\section*{Acknowledgements}

The research leading to these results has received funding from the European Commission under the grant agreement eHeroes (project n. 284461, www.eheroes.eu).

\end{document}